%% file: main.tex
\documentclass[sigconf]{acmart}

\usepackage{textcomp}
\usepackage{stfloats}
\usepackage{url}
\usepackage{verbatim}
\usepackage{amsthm}
\usepackage{subfigure}
\usepackage{makecell}
\usepackage{multirow}
\usepackage{enumitem}
\usepackage{subcaption}

\usepackage{color, colortbl}

\definecolor{Gray}{rgb}{0.9, 0.9, 0.9}

\usepackage{float}

\AtBeginDocument{%
  }

\copyrightyear{2025}
\acmYear{2025}
\setcopyright{acmlicensed}\acmConference[SIGIR '25]{The 48th International ACM SIGIR Conference on Research and Development in Information Retrieval}{July 13--17, 2025}{Padua, Italy}
\acmBooktitle{The 48th International ACM SIGIR Conference on Research and Development in Information Retrieval (SIGIR 2025), July 13--17, 2025, Padua, Italy}
\acmDOI{}
\acmISBN{}

\setcopyright{acmlicensed}
\copyrightyear{2025}
\acmYear{2025}
\acmDOI{XXXXXXX.XXXXXXX}
\begin{document}

\title[WeightedGCL]{Squeeze and Excitation: A Weighted Graph Contrastive Learning for Collaborative Filtering}


\settopmatter{authorsperrow=4}

\author{Zheyu Chen}
\affiliation{%
  \institution{\small{The Hong Kong Polytechnic University}}
  \city{Hong Kong SAR}
  \country{China}}
\email{zheyu.chen@connect.polyu.hk}

\author{Jinfeng Xu}
\authornote{Corresponding author}
\affiliation{%
  \institution{\small{The University of Hong Kong}}
  \city{Hong Kong SAR}
  \country{China}}
\email{jinfeng@connect.hku.hk}

\author{Yutong Wei}
\affiliation{%
  \institution{\small{Central South University}}
  \city{Changsha}
  \country{China}}
\email{8207220511@csu.edu.cn}

\author{Ziyue Peng}
\affiliation{%
  \institution{\small{University of Macau}}
  \city{Macao SAR}
  \country{China}}
\email{dc22951@um.edu.mo}

\renewcommand{\shortauthors}{Zheyu Chen et al.}

\begin{abstract}
Contrastive Learning (CL) has recently emerged as a powerful technique in recommendation systems, particularly for its capability to harness self-supervised signals from perturbed views to mitigate the persistent challenge of data sparsity. The process of constructing perturbed views of the user-item bipartite graph and performing contrastive learning between perturbed views in a graph convolutional network (GCN) is called graph contrastive learning (GCL), which aims to enhance the robustness of representation learning. Although existing GCL-based models are effective, the weight assignment method for perturbed views has not been fully explored. A critical problem in existing GCL-based models is the irrational allocation of feature attention. This problem limits the model's ability to effectively leverage crucial features, resulting in suboptimal performance. To address this, we propose a Weighted Graph Contrastive Learning framework (WeightedGCL). Specifically, WeightedGCL applies a robust perturbation strategy, which perturbs only the view of the final GCN layer. In addition, WeightedGCL incorporates a squeeze and excitation network (SENet) to dynamically weight the features of the perturbed views. Our WeightedGCL strengthens the model's focus on crucial features and reduces the impact of less relevant information. Extensive experiments on widely used datasets demonstrate that our WeightedGCL achieves significant accuracy improvements compared to competitive baselines.
\end{abstract}

% <concept_id>10002951.10003317.10003347.10003350</concept_id>
% <concept_desc>Information systems~Recommender systems</concept_desc>



\maketitle

\input{Tex/1.Introduction}
\input{Tex/2.Methodology}

\input{Tex/3.Experiment}
\input{Tex/4.Conclusion}

% \newpage

% \begin{acks}
% xxxxxx
% \end{acks}

% \balance
% \bibliographystyle{ACM-Reference-Format}
% \bibliography{reference}

\input{arxiv.bbl}

\end{document}

%% file: Tex/1.Introduction.tex
\section{Introduction}
To alleviate the problem of information overload on the web, recommender systems have become an important tool \cite{xu2024improving, xu2024mentor, he2017neural, xu2024aligngroup}, which commonly adopt collaborative filtering (CF) to capture the complex relations between users and items.
Attribute to the natural bipartite graph structure of user-item interactions, many existing works have achieved advanced performance by effectively leveraging graph-based recommendation models \cite{he2020lightgcn, wang2019neural}. For example, NGCF \cite{wang2019neural} integrates a graph convolutional network (GCN) framework into recommender systems, maintaining both feature transformation and non-linear operations.
In contrast, LightGCN \cite{he2020lightgcn} raises that these components are unnecessary for recommendation tasks, proposing a lightweight GCN model instead. Numerous subsequent graph-based models \cite{zhou2023layer, wei2019mmgcn, yu2023xsimgcl} have pushed the boundaries of graph-based recommendation systems, offering improved scalability, accuracy, and robustness.

In recent years, contrastive learning (CL) has seen great development in useless representation learning. In the recommendation field, considering the persistent challenge of data sparsity, CL has proven to be a powerful self-supervised learning approach for leveraging unlabeled data.

Recent studies \cite{wu2021self, chen2025don, cailightgcl, yu2023xsimgcl} have demonstrated the effectiveness of combining GCN and CL, termed graph contrastive learning (GCL), to improve recommendation performance. GCL typically involves constructing perturbed views of the user-item bipartite graph and applying contrastive learning between these views.

Nevertheless, these existing models \cite{wu2021self, yu2022graph} often apply uniform weights for all features within each perturbed view, neglecting the varying significance of different features. The equal weight strategy will limit the model's ability to effectively leverage crucial features, making it pay too much attention to less relevant information. This raises a crucial question: \textit{\textbf{How to allocate attention to different features dynamically?}}

To address this, we propose a tailored and novel framework named \textbf{Weighted} \textbf{G}raph \textbf{C}ontrastive \textbf{L}earning (\textbf{WeightedGCL})\footnote{Code is available at: \href{https://github.com/Zheyu-Chen/WeightedGCL}{https://github.com/Zheyu-Chen/WeightedGCL}.}. Specifically, WeightedGCL applies a robust perturbation strategy, which only perturbs the final GCN layer's view. In addition, WeightedGCL incorporates a squeeze and excitation network (SENet) \cite{hu2018squeeze} to dynamically assign weights to the features of these perturbed views. Our WeightedGCL enhances the attention on the crucial features by assigning greater weight to them. At the same time, it reduces the attention to less relevant information, ensuring that the model's performance is not degraded by this information. In a nutshell, the contributions of our work are as follows:
\vspace{-1mm}
\begin{itemize}[leftmargin=*]
\item We identify the limitation in existing GCL-based frameworks, in which they assign equal weight to all features and consequently propose a dynamic feature weighting solution.
\item We propose a Weighted Graph Contrastive Learning framework, which incorporating the robust perturbation to keep views stable and incorporating the SENet to assign weights dynamically.
\item Experiment results on three public datasets demonstrate the effectiveness of our WeightedGCL.
\end{itemize}
\vspace{-3mm}

%% file: Tex/2.Methodology.tex
\section{Methodology}
\begin{figure}
    \centering
    \includegraphics[width=1\linewidth]{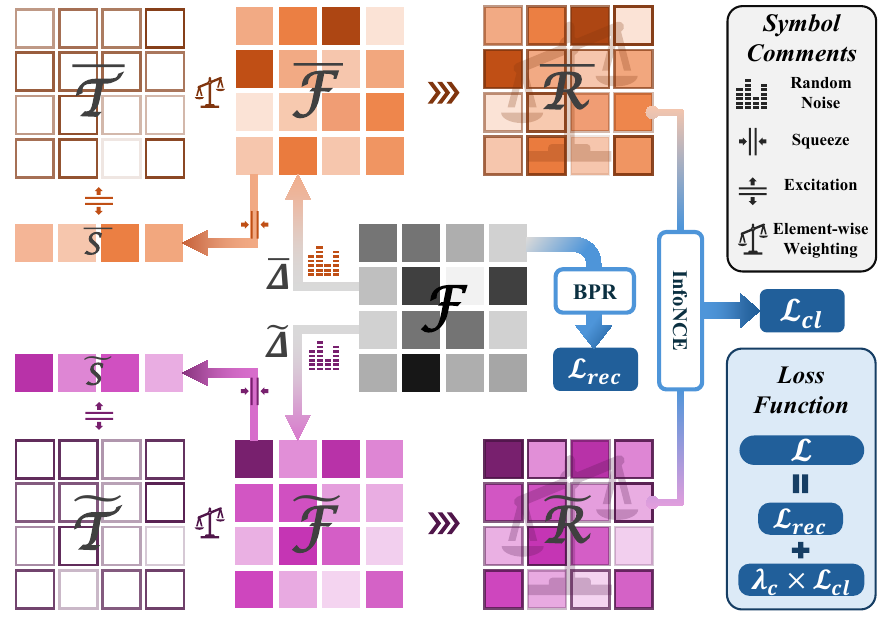}
    \vskip -0.15in
    \caption{Overall architecture of our WeightedGCL.}
    \label{fig:overview}
    \vskip -0.15in
\end{figure}

In contrastive learning, the construction of contrastive views is crucial. The previous work \cite{yu2023xsimgcl} on GCN-based perturbation applies perturbations across all layers of the network. However, this work introduces noise at earlier layers, which can destabilize the representation learning process. To address this, we propose a robust perturbation strategy that only applies perturbations to the final layer of the GCN. This targeted perturbation avoids the negative impact of early-stage noise, thereby maintaining more stable and meaningful learned representations. Due to the inherent randomness in perturbed views, directly learning the weight matrix proves challenging. To address this challenge, we introduce a SENet, which includes two parts: the squeeze network and the excitation network. The former reduces the dimensions of each perturbed view into a summary statistics matrix, and then the latter transforms the matrix back to the original dimensions, which assign the feature weights dynamically. Therefore, we propose a dynamic feature weighting solution by combining SENet and GCL, named WeightedGCL. Our WeightedGCL enhances the focus on crucial features while effectively mitigating the impact of less relevant information, and achieving performance improvement. The overall architecture of WeightedGCL is depicted in Figure~\ref{fig:overview}. 

\subsection{Preliminary}
Let $ \mathcal{U} $ denotes the set of users and $ \mathcal{I} $ denotes the set of items. The $ L $ represents the number of layers in the GCN. The set of all nodes, including both users and items, is denoted as $ \mathcal{N} = \mathcal{U} \cup \mathcal{I} $. Consider a node $ n $ within the set $ \mathcal{N} $, whose representation is $ e_n \in \mathbb{R}^{d \times 1} $ in the view $ \mathcal{E} \in \mathbb{R}^{d \times |\mathcal{N}|} $, where $ d $ denotes the feature dimension.

\subsection{Feature Encoder}
GCNs refine node representations by aggregating messages from their neighboring nodes. This process can be formalized as follows:
\vskip -0.15in
\begin{equation}
\begin{aligned}
{e}_u^{(l)}&=\operatorname{Aggr}^{(l)}(\{{e}_i^{(l-1)}: i \in \mathcal{N}_u\}), \\
{e}_i^{(l)}&=\operatorname{Aggr}^{(l)}(\{{e}_u^{(l-1)}: u \in \mathcal{N}_i\}),
\end{aligned}
\end{equation}
\vskip -0.05in
\noindent where $\mathcal{N}_u$ and $\mathcal{N}_i$ denote the neighborhood set of nodes $u$ and $i$, respectively, and $l$ denotes the $l$-th layer of GNNs. The node representation aggregation for the entire embeddings $\mathcal{{F}}$ can be formulated as follows:
\vskip -0.15in
\begin{equation}
\mathcal{{F}}=\frac{1}{L+1}(\mathcal{E}^{(0)}+\mathcal{E}^{(1)}+...+\mathcal{E}^{(L-1)}+{\mathcal{E}}^{(L)}),
\end{equation}
\vskip -0.05in
\noindent where $\mathcal{E}^{(l)}$ denotes the view of nodes in $l$-th layer.

\subsection{Robust Perturbation}
We initially construct contrastive views by the robust perturbation strategy, which adds a random noise matrix into the view of the final layer in GCN. Formally:
\vskip -0.15in
\begin{equation}
\bar{\mathcal{E}}^{(L)} = {\mathcal{E}}^{(L)} + \bar{\Delta}, \quad
\tilde{\mathcal{E}}^{(L)} = {\mathcal{E}}^{(L)} + \tilde{\Delta},
\end{equation}
\vskip -0.15in
\begin{equation}
\begin{aligned}
\mathcal{\bar{F}}&=\frac{1}{L+1}(\mathcal{E}^{(0)}+\mathcal{E}^{(1)}+...+\mathcal{E}^{(L-1)}+\bar{\mathcal{E}}^{(L)}),\\
\mathcal{\tilde{F}}&=\frac{1}{L+1}(\mathcal{E}^{(0)}+\mathcal{E}^{(1)}+...+\mathcal{E}^{(L-1)}+\tilde{\mathcal{E}}^{(L)}),
\end{aligned}
\end{equation}

\noindent where $\mathcal{\bar{F}}$ and $\mathcal{\tilde{F}}$ denote final representations of two perturbed views. Meanwhile, the $\bar{\Delta}$ and $\tilde{\Delta}$ $\in \mathbb{R}^{d \times |\mathcal{N}|} \sim U(0,1)$ are random noise matrixs. Only perturbing the final layer will maintain the view stability of other front layers. We will conduct an ablation study in Section ~\ref{sec:ablation study} to demonstrate the effectiveness of our strategy.

\subsection{Squeeze Network}
The squeeze network adopts average pooling, which is beneficial for retaining feature information, to reduce the dimension of the entire perturbed views $\ddot{\mathcal{F}}$ into a summary statistics matrices $\mathcal{\bar{S}}/\mathcal{\tilde{S}} \in \mathbb{R}^{1 \times |\mathcal{N}|}$, formally:
\vskip -0.25in
\begin{equation}
\begin{aligned}
\operatorname{Squeeze}(\ddot{\mathcal{F}}) 
% = Con({z}_n  \mid n \in \mathcal{U} \cup  \mathcal{I}) \\ &
= \operatorname{Con}(\frac{1}{d} \sum_{k=1}^{d} \ddot{f}_{n}^{k} \mid n \in \mathcal{N}),
\end{aligned}
\end{equation}
\vskip -0.10in
\begin{equation}
\mathcal{\bar{S}} = \operatorname{Squeeze}(\mathcal{\bar{F}}), \quad \mathcal{\tilde{S}} = \operatorname{Squeeze}(\mathcal{\tilde{F}}),
\label{eq:3}
\end{equation}
\vskip -0.05in
\noindent where $k$ represents the $k$-th feature dimension of perturbed node representation $\ddot{f}_n$, and $\frac{1}{d} \sum_{k=1}^{d} \ddot{f}_{n}^{k}$ is the summary statistics matrix of node $n$.

\subsection{Excitation Network}
The summary statistics matrix is then expanded back to the original dimensions using the excitation network, formally:
\vskip -0.15in
\begin{equation}
\operatorname{Excitation}(\mathcal{S}) = \sigma(W_K (...(W_1 \cdot \mathcal{S} + b_1)...) + b_K),
\label{eq:5}
\end{equation}
\vskip -0.15in
\begin{equation}
\mathcal{\bar{T}} = \operatorname{Excitation}(\mathcal{\bar{S}}), \quad \mathcal{\tilde{T}} = \operatorname{Excitation}(\mathcal{\tilde{S}}),
\label{eq:6}
\end{equation}
\vskip -0.05in
\noindent where the resulting matrix $\mathcal{\bar{T}}/\mathcal{\tilde{T}} \in \mathbb{R}^{d \times |\mathcal{N}|}$ serves as the weight matrix corresponding to the perturbed views $ \bar{\mathcal{F}}/\tilde{\mathcal{F}}$ , and $\sigma$ denotes the sigmoid function. $K$ is the feedforward network layer number in Eq.~\ref{eq:5}. The excitation network employs a multi-layer architecture, gradually ascending the dimension according to an equal scale $s=\sqrt[K]{d} $ until the original dimensions are restored, and the $W_1 \in \mathbb{R}^{s \times 1}$, ...,  $W_K \in \mathbb{R}^{d \times s^{(K-1)}}$ are the weight matrices for the linear layers. Maintaining a constant ratio facilitates the simplification of the training process. This design results in a more precise generation of the weight matrix, thereby improving the robustness. 

\subsection{Recalibration}
Eventually, the weighted views $\mathcal{\bar{R}} / \mathcal{\tilde{R}} \in  \mathbb{R}^{d \times |\mathcal{N}|}$ are obtained by scaling the perturbed views ${\mathcal{\bar{F}}} / \mathcal{\tilde{F}}$ with dynamic weight matrix $\mathcal{\bar{T}}/\mathcal{\tilde{T}}$, formally:
\vskip -0.2in
\begin{equation}
\mathcal{\bar{R}} = \mathcal{\bar{T}} \odot {\mathcal{\bar{F}}}, \quad \mathcal{\tilde{R}} = \mathcal{\tilde{T}} \odot {\mathcal{\tilde{F}}},
\end{equation}
\vskip -0.05in
\noindent where the $\odot$ represents element-wise multiplication.

\subsection{Contrastive Learning}
We adopt the InfoNCE \cite{oord2018representation} loss function to perform contrastive learning between two perturbed views. Formally, the loss function is defined as follows:
\vskip -0.2in
\begin{equation}
\begin{aligned}
\mathcal{L}_{cl} &=-\sum_{u \in \mathcal{U}}\log \frac{e^{(\bar{r}_{u}^{\top} \tilde{r}_{u} / \tau)}}{\sum_{{u}^{'} \in \mathcal{U}} e^{(\bar{r}_{u}^{\top} \tilde{r}_{{u}^{'}}/ \tau)}}- \sum_{i \in \mathcal{I}}\log \frac{e^{(\bar{r}_{i}^{\top} \tilde{r}_{i} / \tau)}}{\sum_{{i}^{'} \in \mathcal{I}}e^{(\bar{r}_{i}^{\top} \tilde{r}_{{i}^{'}}/ \tau)}},
\end{aligned}
\end{equation}
\vskip -0.05in
\noindent where $\bar{r}_{{u}}$ and $\tilde{r}_{{u}/{u}^{'}}$ are representation of user ${u}/{u}^{'}$ in contrastive views $\bar{\mathcal{R}}$ and ${\tilde{\mathcal{R}}}$. Besides, $\bar{r}_{{i}}$ and $\tilde{r}_{{i}/{i}^{'}}$ are representation of item ${i}/{i}^{'}$in contrastive views $\bar{\mathcal{R}}$ and ${\tilde{\mathcal{R}}}$. $\tau$ is the temperature hyper-parameter.

\begin{table*}[!ht]
\centering
\small
\tabcolsep=0.1cm
\caption{Performance comparison of baselines,  WeightedGCL and variants of WeightedGCL in terms of Recall@K(R@K) and NDCG@K(N@K). The superscript $^*$ indicates the improvement is statistically significant where the $p$-value is less than 0.01.}
\vskip -0.15in
\label{tab:comparison results}
% \resizebox{0.75\linewidth}{!}{
\begin{tabular}{ccccccccccccc}
\toprule
\multirow{2.5}{*}{\textbf{Model}}  & \multicolumn{4}{c}{\textbf{Amazon}} & \multicolumn{4}{c}{\textbf{Pinterest}} & \multicolumn{4}{c}{\textbf{Alibaba}}\\ \cmidrule(lr){2-5} \cmidrule(lr){6-9}\cmidrule(lr){10-13}& R@10   & {R@20}   & N@10   & {N@20}   & R@10   & {R@20}   & N@10   & N@20    & R@10   & {R@20}   & N@10   &N@20    \\ \midrule
{MF-BPR}               & 0.0607 & {0.0956} & 0.0430 & {0.0537} & 0.0855 & {0.1409} & 0.0537 & 0.0708  & 0.0303& 0.0467& 0.0161&0.0203\\
{NGCF}                 & 0.0617 & {0.0978} & 0.0427 & {0.0537} & 0.0870 & {0.1428} & 0.0545 & 0.0721  & 0.0382& 0.0615& 0.0198&0.0257\\
{LightGCN}             & 0.0797 & {0.1206} & 0.0565 & {0.0689} & 0.1000 & {0.1621} & 0.0635 & 0.0830  & 0.0457& 0.0692& 0.0246&0.0299\\
{UltraGCN}             & 0.0760 & {0.1155} & 0.0540 & {0.0643} & 0.0967 & {0.1588} & 0.0613 & 0.0808  & 0.0411& 0.0644& 0.0227&0.0276\\
{LayerGCN}             & 0.0877 & {0.1291} & 0.0647 & {0.0760} & 0.1004 & {0.1620} & 0.0635 & 0.0826  & 0.0448& 0.0680& 0.0238&0.0285\\
{FKAN-GCF}             & 0.0838 & {0.1265} & 0.0602 & {0.0732} & 0.1003 & {0.1614} & 0.0633 &  0.0827  & 0.0441& 0.0681& 0.0240&0.0290\\ 
\midrule
{SGL}                  & 0.0898 & {0.1331} & 0.0645 & {0.0777} & \underline{0.1080} & \underline{0.1704} & 0.0701 & \underline{0.0897}  & 0.0461& 0.0692& 0.0248&0.0307\\
 {NCL}& 0.0933& 0.1381& 0.0679& 0.0815 & 0.1033& 0.1609& 0.0666& 0.0833& 0.0477& 0.0713& 0.0259&\underline{0.0319}\\
{SimGCL}               & \underline{0.0963} & \underline{0.1336} & \underline{0.0718} & \underline{0.0832} & 0.1051 & {0.1576} & \underline{0.0705} & 0.0871  & 0.0474& 0.0691& \underline{0.0262}&0.0317\\
{LightGCL}             & 0.0820 & {0.1278} & 0.0589 & {0.0724} & 0.0881 & {0.1322} & 0.0534 & 0.0673  & 0.0459& 0.0716& 0.0239&0.0305\\
{DCCF}                 & {0.0903} & {0.1307} & 0.0655 & {0.0781} & 0.1040 & {0.1613} & 0.0661 & 0.0828  & \underline{0.0490}& \underline{0.0729}& 0.0257&0.0311\\
 {RecDCL}& 0.0927& 0.1345& 0.0652& 0.0780& 0.1021& 0.1619& 0.0663& 0.0839& 0.0521& 0.0768& 0.0273&0.0338\\
 {BIGCF}& 0.0948& 0.1341& 0.0692& 0.0810& 0.1040& 0.1619& 0.0680& 0.0864& 0.0502& 0.0744& 0.0266&0.0322\\ \midrule
 WGCL-all pert.& 0.0983& 0.1378& 0.0733& 0.0853& 0.1163& 0.1768& 0.0755& 0.0961& 0.0560& 0.0831& 0.0305&0.0374\\
 WGCL-w/o pert.& 0.0098& 0.0159& 0.0062& 0.0080& 0.0103& 0.0168& 0.0066& 0.0086& 0.0102& 0.0167& 0.0050&0.0067\\ 
\textbf{WeightedGCL}          & \textbf{0.0996$^*$} & \textbf{0.1396$^*$}& \textbf{0.0741$^*$} &  \textbf{0.0862$^*$} & \textbf{0.1167$^*$} & \textbf{0.1793$^*$} & \textbf{0.0764$^*$}& \textbf{0.0961$^*$}  & \textbf{0.0596$^*$}& \textbf{0.0879$^*$}& \textbf{0.0326$^*$}&\textbf{0.0397$^*$}\\ 
\textit{Improv.} &    3.43\%    & {4.49\%}&    3.20\%    & {3.61\%}       &    8.06\%& {5.22\%}&   7.81\%&    7.13\%& 21.63\%& 20.58\%& 24.42\%& 24.45\%\\  \bottomrule
\end{tabular}
% }
\end{table*}
 
\subsection{Optimization}
We utilize LightGCN \cite{he2020lightgcn} as the backbone and adopt the Bayesian Personalized Ranking (BPR) loss \cite{rendle2009bpr} as our primary optimization objective. The BPR loss is designed to enhance the distinction between predicted preferences for positive and negative items in each triplet  $(u, p, n) \in \mathcal{D}$, where $\mathcal{D}$ is the training dataset. The positive item $p$ is an item with which user $u$ has interacted, while the negative item $n$ is randomly selected from the items not interacted with user $u$. The BPR loss function is formally defined as:
\vskip -0.1in
\begin{equation} 
    \mathcal{L}_{rec} = \sum_{(u, p, n) \in \mathcal{D}} - \log(\sigma(y_{u,p} - y_{u,n})) + \lambda \cdot \|\mathbf{\Theta}\|^2_2, 
\end{equation}
\vskip -0.05in
\noindent where $\sigma$ denotes the sigmoid function, $\lambda$ controls the $L_2$ regularization strength, and $\mathbf{\Theta}$ denotes model parameters. The $y_{u,p}$ and $y_{u,n}$ are the ratings of user $u$ to the positive item $p$ and negative item $n$, which calculated by $r_u^{\top}r_p$ and $r_u^{\top}r_n$. Ultimately, the total loss function is:
\vskip -0.15in
\begin{equation}
\mathcal{L} = \mathcal{L}_{rec} + \lambda_{c} \cdot \mathcal{L}_{cl},
\end{equation}
\vskip -0.05in
\noindent where $\lambda_{c}$ is the balancing hyper-parameter.

%% file: Tex/3.Experiment.tex
\section{Experiment}
In this section, we conduct extensive experiments on three real-world datasets to evaluate WeightedGCL, addressing the following research questions: 

\noindent\textbf{RQ1}: How does the performance of our WeightedGCL compare to advanced recommender systems? 

\noindent\textbf{RQ2}: How does our Robust Perturbation component influence the overall performance of our WeightedGCL?

\noindent\textbf{RQ3}: How do different hyper-parameters influence the performance of our WeightedGCL?

\subsection{Experimental Settings}
\subsubsection{Datasets. }
To evaluate our WeightedGCL in the recommendation task, we conduct extensive experiments on three widely used datasets: Amazon Books (Amazon) \cite{mcauley2015image}, Pinterest \cite{he2017neural} and Alibaba. Details can be found in Table~\ref{tab:dataset_statistics}. These datasets offer a set of user-item interactions, and we use the ratio 8:1:1 for each dataset to randomly split the data for training, validation, and testing.

\begin{table}[!ht]
    \centering
    \small
\caption{Statistics of datasets.}
\vskip -0.15in
\label{tab:dataset_statistics}
    \begin{tabular}{ccccc}
         \toprule
         \textbf{Datasets}&  \textbf{\#Users}&  \textbf{\#Items}&  \textbf{\#Interactions}& \textbf{Sparsity}\\
         \midrule
         \textbf{Amazon} & 58,144 & 58,051 & 2,517,437 & 99.925\%\\
         \textbf{Pinterest} & 55,188 & 9,912 & 1,445,622 & 99.736\%\\
         \textbf{Alibaba} & 300,001 & 81,615 & 1,607,813 & 99.993\%\\
         \bottomrule
    \end{tabular}
\vskip -0.1in
\end{table}

To ensure the quality of the data, we employ the 15-core setting for Amazon, which ensures a minimum of 15 interactions between users and items. For the Pinterest datasets, users and items with less than five interactions are filtered out. Due to the high sparsity of the Alibaba dataset, we choose to retain all interaction data.

\subsubsection{Baselines. }To verify the effectiveness of WeightGCL, we select one matrix factorization model \textbf{MF-BPR} \cite{rendle2009bpr}, five advanced GCN-based models (\textbf{NGCF} \cite{wang2019neural}, \textbf{LightGCN} \cite{he2020lightgcn}, \textbf{UltraGCN} \cite{mao2021ultragcn},  \textbf{LayerGCN} \cite{zhou2023layer} and \textbf{FKAN-GCF} \cite{xu2024fourierkan}), and seven state-of-the-art GCL-based models (\textbf{SGL}\cite{wu2021self}, \textbf{NCL} \cite{lin2022improving}, \textbf{SimGCL} \cite{yu2022graph}, \textbf{LightGCL} \cite{cailightgcl}, \textbf{DCCF} \cite{ren2023disentangled}, \textbf{RecDCL} \cite{zhang2024recdcl} and \textbf{BIGCF} \cite{zhang2024exploring}) for comparison.

\subsubsection{Hyper-parameters. }
To ensure a fair comparison, we initially refer to the optimal hyper-parameter settings as reported in the original papers of the baseline models. Subsequently, we fine-tune all hyper-parameters of the baselines using grid search. For the general settings across all baselines, we apply Xavier initialization \cite{glorot2010understanding} to all embeddings. The embedding size is 64, and the batch size is 4096. All models are optimized using the Adam optimizer \cite{kingma2014adam} with a learning rate of 1e-4. We perform a grid search on the balancing hyper-parameter $\lambda_c$ in \{1e-1, 1e-2, 1e-3\}, temperature hyper-parameter $\tau$ in \{0.2, 0.4, 0.6, 0.8\}, and the layer number $L$ of GCN in \{2, 3, 4\}. Empirically, we set the $L_2$ regularization parameter to 1e-4 for Amazon and Pinterest and 1e-5 for Alibaba. To avoid the over-fitting problem, we set 30 as the early stopping epoch number. Moreover, we set the excitation network boosting granularity as a hyper-parameter, containing 1-4 levels of granularity, and detailed description and analysis are in Section \ref{sec:hyper-parameter sensitivity study}.

\subsubsection{Evaluation Protocols.}
For fairness, we follow the settings of previous works \cite{he2017neural,wang2019neural, xu2025survey} by adopting two widely-used evaluation protocols for top-$K$ recommendations: Recall@K and NDCG@K\cite{jarvelin2002cumulated}. We report the average metrics across all users in the test dataset, evaluating performance at both $K=10$ and $20$.

\subsection{Performance Comparison (RQ1)}
Detailed experiment results are shown in Table~\ref{tab:comparison results}. The optimal results are highlighted in bold, while the suboptimal results are indicated with underlines. Based on these results, we observed that: our WeightedGCL outperforms the strongest baselines, achieving 4.49\%(R@20), 3.61\%(N@20) improvement on the Amazon dataset, achieving 5.22\%(R@20), 7.13\%(N@20) improvement on the Pinterest dataset, and achieving 20.58\%(R@20), 24.45\%(N@20) improvement on the Alibaba dataset, which demonstrates the effectiveness of WeightedGCL. Moreover, we can identify that most GCL-based models outperform traditional models, which is a common trend. Our WeightedGCL outperforms all baselines in three datasets for various metrics.

\subsection{Ablation Study (RQ2)}
\label{sec:ablation study}
To validate the effectiveness of our robust perturbation, we design the following variants: 

\noindent \textbf{WGCL-w/o pert.} For this variant, we directly removed our robust perturbation component. 

\noindent \textbf{WGCL-all pert.} For this variant, we utilize all layer's perturbation to replace our robust perturbation component.

Table~\ref{tab:comparison results} also demonstrates the significance of perturbation strategies. This ablation study shows that the method's performance without perturbation has a sharp performance degradation, resulting in a completely unusable model, which indicates that the perturbation operation is essential for contrastive learning. Our robust perturbation outperforms other variants, which indicates the robust perturbation component plays a critical role in improving model representation power and model robustness.

\subsection{Hyper-parameter Sensitivity Study (RQ3)}
\label{sec:hyper-parameter sensitivity study}
To evaluate the hyper-parameter sensitivity of WeightedGCL, we test its performance on three datasets under varying hyper-parameters. Table~\ref{tab:excitation} highlights the impact of various excitation strategies. We denote different ascending granularity as W-G1, W-G2, W-G3, and W-G4, which correspond to varying numbers of layers in the FFN within the excitation network. The larger the number, the finer the granularity, and the more layers.

\begin{table}[!h]
\vskip -0.1in
    \centering
    \small
    \tabcolsep=0.15cm
\caption{Excitation strategies analysis.}
\vskip -0.15in
\label{tab:excitation}
\begin{tabular}{ccccccc}
\toprule
\multirow{2.5}{*}{\textbf{Variant}}                                 &\multicolumn{2}{c}{Amazon}&\multicolumn{2}{c}{Pinterest} & \multicolumn{2}{c}{Alibaba}\\ \cmidrule(lr){2-3} \cmidrule(lr){4-5} \cmidrule(lr){6-7}
                                  & R@20   & N@20   & R@20   & N@20    & R@20   &N@20    \\
\midrule
\multicolumn{1}{c}{W-G1}                    & 0.1314 & 0.0807 & 0.1713 & 0.0931  & 0.0772& 0.0359\\
\multicolumn{1}{c}{W-G2}                    & 0.1392& 0.0859 & 0.1764 & 0.0953  & 0.0793& 0.0363\\
\multicolumn{1}{c}{W-G3}                     & 0.1391 & 0.0860 & \textbf{0.1793} & \textbf{0.0961} &\textbf{0.0879} &\textbf{0.0397}\\
\multicolumn{1}{c}{W-G4}                     & \textbf{0.1396}& \textbf{0.0862} & 0.1790 & 0.0956 & 0.0826& 0.0371\\
\bottomrule
\end{tabular}
\vskip -0.1in
\end{table}

After analyzing the results in Table~\ref{tab:excitation}, we found that our framework performs best at the fourth granularity for the Amazon dataset, and the third granularity is the best for the Pinterest and Alibaba datasets. Although the fourth granularity is not the best choice for the latter two datasets, it is still higher than the performance at the first and second granularities. These observations illustrate two points: from the overall trend, the performance of our incentive network is improving with the increase of granularity; but too high granularity makes it difficult to train due to the increased complexity of the model, which will slightly reduce the performance in some cases. This ablation study demonstrates that the choice of excitation strategies to some extent influences our framework's performance.

\begin{figure}[h]
\vskip -0.1in
    \centering
        \includegraphics[width=1\linewidth]{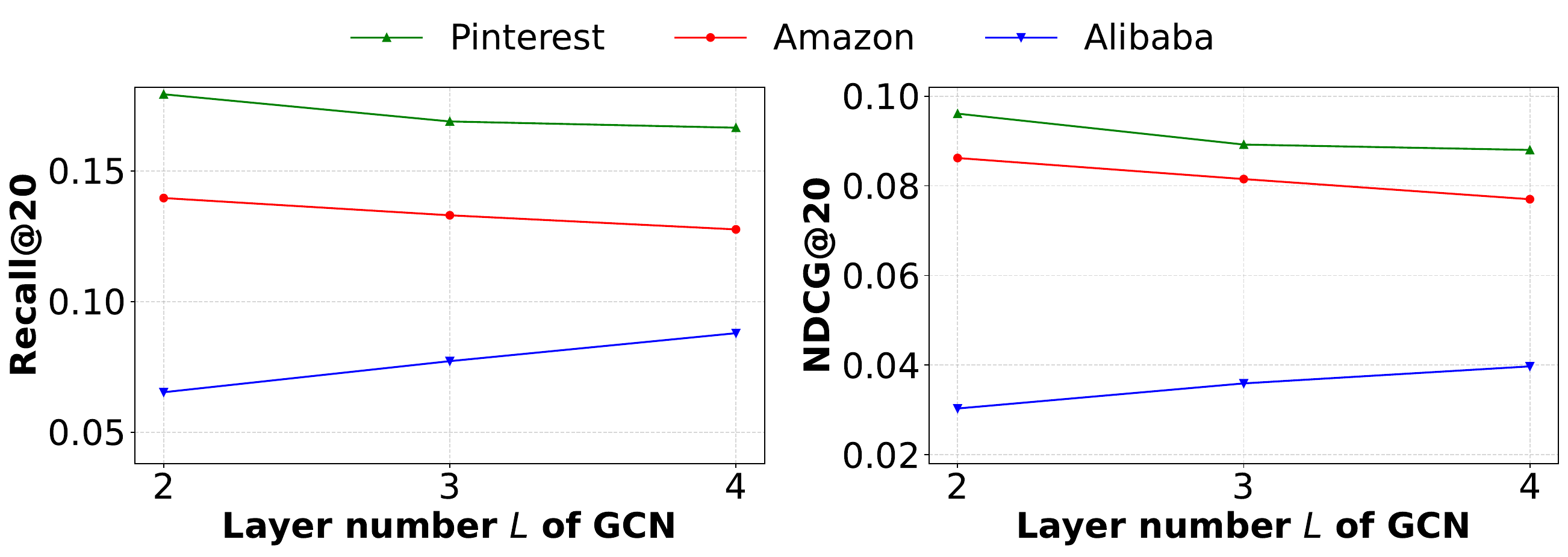}
    \vskip -0.15in
    \caption{Performance of our WeightedGCL with respect to different layer number $L$ of GCN.}
    \label{fig:layer number}     
    \vskip -0.1in
\end{figure}

\vskip -0.15in

\begin{figure}[h]
\vskip -0.1in
    \centering
    \subfigure {
        \label{fig:H_Amazon}
        \includegraphics[width=0.33\linewidth]{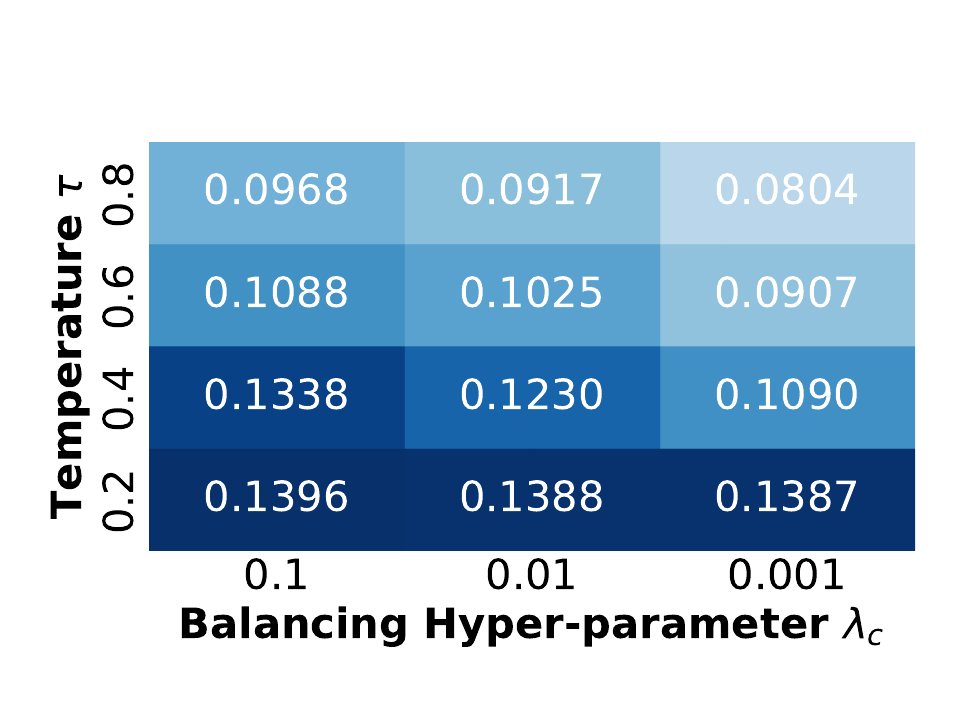}
        } \hspace{-0.15in}
    \subfigure {
        \label{fig:H_Pinterest}
        \includegraphics[width=0.33\linewidth]{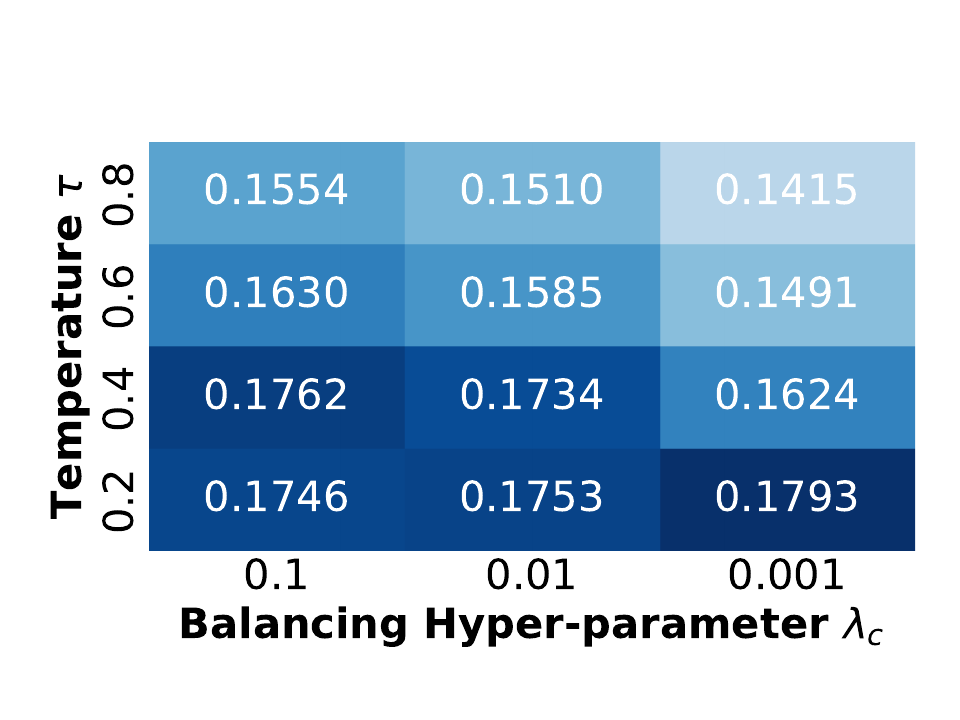}
        } \hspace{-0.15in}
    \subfigure {
        \label{fig:H_Alibaba}
        \includegraphics[width=0.33\linewidth]{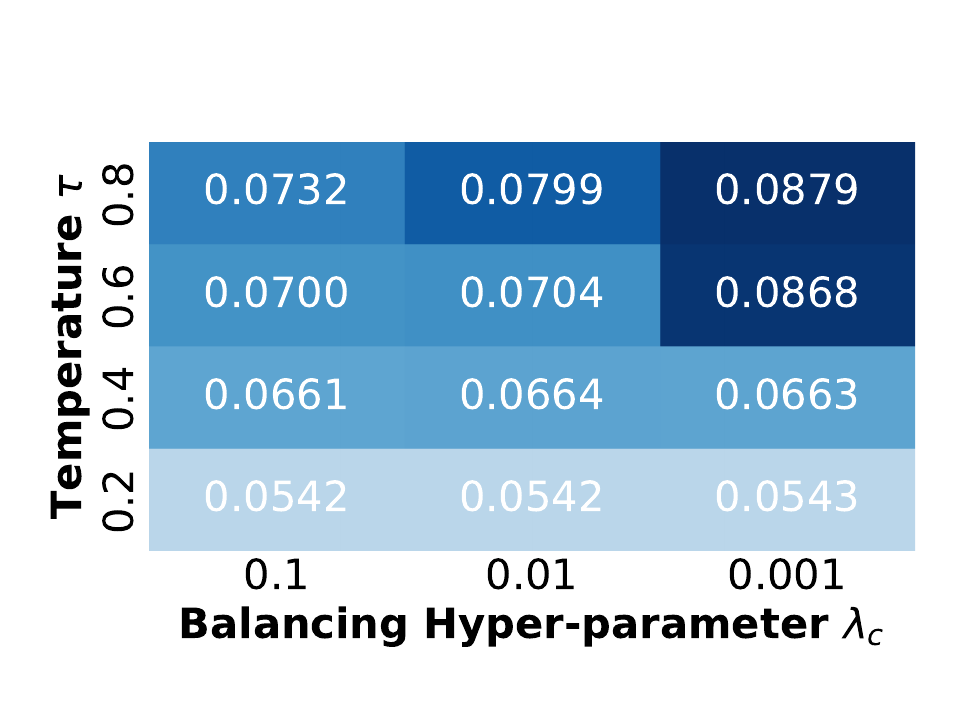}
        }
    \vskip -0.25in
    \caption{Performance of our WeightedGCL with respect to different hyper-parameter pairs $(\lambda_{c}, \tau)$ in terms of Recall@20.}
    \label{fig:heatmap}    
\vskip -0.1in
\end{figure}

As shown in the Figure~\ref{fig:layer number}, for the Amazon and Pinterest datasets, the optimal layer number $L$ is 2, and for the Alibaba dataset, the optimal number is 3. Additionally, Figure~\ref{fig:heatmap} reveals that for the Amazon dataset, the optimal $(\lambda_{c}, \tau)$ pair is (1e-1, 0.2); for Pinterest dataset, the optimal pair is (1e-3, 0.2); and for the Alibaba dataset, (1e-3, 0.8) is the optimal pair. The optimal temperature hyper-parameter $\tau$ for the Alibaba dataset differs from the other two datasets, attributed to its sparse user-item interactions and the huge number of users. Note that being flexible in choosing the value of hyper-parameters will allow us to adopt our framework to multiple datasets. 

%% file: Tex/4.Conclusion.tex
\section{Conclusion}
In this paper, we propose WeightedGCL, a novel model that incorporates a tailored robust perturbation strategy and SENet with GCL. Existing GCL-based models assign equal weights for all features within each perturbed view, which limits the model's ability to effectively leverage crucial features. Our WeightedGCL enhances the attention on the crucial features by assigning greater weight to them and reduces the attention to less relevant information. Our experiments on three widely used datasets show that WeightedGCL achieves significant performance improvements compared to existing models. The improvement demonstrates the effectiveness of WeightedGCL and its potential to advance the development of recommendation systems.